# High-Frequency Microstrip Cross Resonators for Circular Polarization EPR Spectroscopy

J. J. Henderson, C. M. Ramsey, H. M. Quddusi and E. del Barco

*Physics Department, University of Central Florida, 4000 Central Florida Boulevard, Orlando, Florida 32816-2385*

## ABSTRACT

In this article we discuss the design and implementation of a novel microstrip resonator which allows for the absolute control of the microwaves polarization degree for frequencies up to 30 GHz. The sensor is composed of two half-wavelength microstrip line resonators, designed to match the 50 Ω impedance of the lines on a high dielectric constant GaAs substrate. The line resonators cross each other perpendicularly through their centers, forming a cross. Microstrip feed lines are coupled through small gaps to three arms of the cross to connect the resonator to the excitation ports. The control of the relative magnitude and phase between the two microwave stimuli at the input ports of each line allows for tuning the degree and type of polarization of the microwave excitation at the center of the cross resonator. The third (output) port is used to measure the transmitted signal, which is crucial to work at low temperatures, where reflections along lengthy coaxial lines mask the signal reflected by the resonator. EPR spectra recorded at low temperature in an $S = 5/2$ molecular magnet system show that 82%-fidelity circular polarization of the microwaves is achieved over the central area of the resonator.

Electron paramagnetic resonance (EPR), often called electron spin resonance (ESR), is a widely used spectroscopy technique for the characterization of magnetic systems of very different nature. In particular, both continuous wave and time-resolved pulse EPR spectroscopy have become a key characterization tool in every chemistry department due, in part, to its ability to provide structural information of organic and inorganic compounds and ongoing chemical and physical processes without interfering with the process itself. Much effort has been directed to extending the range of applicability of this technique to higher frequencies, shorter pulse times, increased sensitivity and/or circular polarized waves. Particularly, controlling the polarization of high frequency microwaves in resonant cavities is of special importance since it allows transitions between different spin states to be selected. The ability to discriminate spin transitions is crucial for spin-based architectures in quantum information and computation applications. A number of spin-based systems have been proposed as potential hardware for quantum computation [1-4]. Circular polarized microwave irradiation can improve the initialization, manipulation and read-out of the qubits by appropriate tuning of the desired transitions. For example, the control of the microwaves polarization could be used to efficiently select different spin transitions for the implementation of the Grover's search algorithm in single-molecule magnets [4]. However, controlling the polarization in high frequency resonant cavities is a technical challenge. Existing approaches implemented in geometrical cavities make use of mechanically adjustable parts to control the polarization degree at resonance [5-11], imposing a limitation for both the arbitrary generation of polarized waves and the integration of polarizable resonators in miniaturized circuitry. In addition, the use of movable mechanical parts to achieve polarization strongly limits the fast generation of pulses with different polarization necessary in quantum information processing.

In this article we present a novel polarizable EPR transmission *microstrip cross resonator*, composed of two half-wavelength microstrip line resonators which allows an in-situ and all-electronic arbitrary control of the polarization of microwaves for frequencies up to 30 GHz. We demonstrate the feasibility of our sensor by measuring the EPR spectra of a sub-millimeter scale single-crystal of magnetically dilute molecular magnets ($S = 5/2$). The results demonstrate the high sensitivity of our sensor, its workability at low temperature, and the achievement of 82%-fidelity circular polarized microwaves, averaged over the central area of the resonator (0.4×0.4 mm$^2$).

Half-wavelength microstrip line resonators have been employed for EPR spectroscopy measurements since 1974 [12,13]. Recently, simultaneous EPR spectroscopy and magnetometry measurements have been demonstrated in a sensor integrating a microstrip resonator and a 2-d electron gas Hall-effect magnetometer on a chip [14]. A typical half-wavelength microstrip resonator is formed by an isolated microstrip transmission line, whose length determines the fundamental resonant mode. The resonator is coupled to one/two ports by means of feeding transmission lines to form a reflection/transmission resonator. Fig. 1 shows the assembly of our polarizable resonator. Two equal length microstrip line resonators designed to work at a frequency range of 5-30 GHz for its fundamental oscillation mode ($L \sim \lambda/2$) are put together forming a square cross. The geometry of the microstrip device (i.e. width of the line, $w$, and thickness of the dielectric substrate, $h$, separating the line from the metallic bottom plate) has been calculated to match the impedance of the microwave coaxial lines (50 Ω) [15]. Note that in our case, where $h < \lambda/2$, the signal transmission corresponds to a quasi-TEM mode in each of the resonator lines. At their fundamental resonant mode, the microwave magnetic field components, $h_{ac}^1$ and $h_{ac}^2$, are maximum and perpendicular to each other at the center of the resonators (see red and blue arrows in Fig. 1). The cross resonator is connected to ports $P_1$ and $P_2$ through feeding lines separated by a *coupling gap*, $g_c$. The gaps are designed to optimally couple the resonator to the feed lines, resulting in a loaded quality factor, $Q_L = Q_0/2$, which is typically about 100 when critically coupled. A third line, connected to port $P_3$, is coupled to an opposite resonator arm through a *transmission gap*, $g_t$ ($>g_c$), which is designed to allow measurements in transmission mode (i.e. $S_{31}$ parameter), without affecting the response of the resonator at resonance [14]. Semi-rigid coaxial lines are used to connect the three ports of the resonator, which is housed at the base of a cryostat, to a 50 GHz Agilent Technologies PNA vector network analyzer. A power splitter, a phase shifter and variable attenuators are used to control the magnitude and phase of the input signals at ports $P_1$ and $P_2$.

Fig. 2 shows the reflection ($S_{11}$) and transmission ($S_{31}$) parameters measured at room temperature (RT) when microwave excitation is sent only to port $P_1$. The data corresponds to a cross resonator with the following geometrical parameters: $L = 4.8$ mm, $g_c = 40$ µm, $g_t = 250$ µm, $w = 480$ µm and $h = 0.6$ mm (see Fig. 1). Two clean resonances (Q ~100) and $f_b = 9.7$ GHz and $f_q = 11.5$ GHz are observed in this resonator. Similar double-resonance response was observed in other resonators designed to work at higher frequencies (not shown). The simulated response of the resonator is also presented in Fig. 2 for comparison (dotted lines) [16], showing excellent agreement with the experimental data. The graphics inset on Fig. 2 represent the current density at each of the resonances, when stimulus is supplied to port $P_1$ (white arrow). At the first resonant frequency, $f_d = 9.7$ GHz (RT), a dipolar resonant excitation develops along one of the arms of the cross resonator, which corresponds to the fundamental resonant excitation mode of an individual microstrip line resonator, which can be estimated as $f_0 \sim nc/(2(L+\Delta L)\sqrt{\varepsilon_r})$, where $n$ is the resonant mode, $c$ the speed of light in vacuum, $L$ the resonator length, $\varepsilon_r$ the effective dielectric constant of the device and $\Delta L$ a correction factor to take into account the effect of the gap [12]. The same excitation occurs along the other arm of the cross resonator when feeding with microwaves of the same frequency in port $P_2$. The second resonance, $f_q = 11.5$ GHz, corresponds to a quadrupole resonant excitation of the whole cross resonator, which develops equally when feeding port $P_2$ with microwaves.

The observed response of the resonator upon excitation of each of the input ports provides the principle of operation of our polarizable resonator. For this, microwave stimuli are applied to both ports $P_1$ and $P_2$. As said above, the relative magnitude and phase between the two microwave signals modulates the response of the resonator. Fig. 3 shows a contour plot of the transmission parameter ($S_{31}$) as a function of the frequency and the phase delay between both signals. Both resonances show a $2\pi$-period modulation.

Let us center our attention into the second resonance first ($f_q = 11.5$ GHz). Interference between the two microwave signals modulates the quadrupole resonant excitation of the whole cross. By adjusting the magnitudes of the signals, one can completely eliminate the quadrupole resonance when the two microwave signals are ±180 degrees out-of-phase (destructive interference). On the contrary, in-phase microwave signals add together (constructive interference) to develop a maximum in the transmission parameter ($S_{31}$). Significantly, the quadrupole resonance can be employed to adjust the relative magnitude and phase between both input signals. This is particularly important when working at low temperature, where long coaxial lines wiring the cryostats strongly affect the microwave power arriving to the resonator input ports.

Most importantly, our results indicate that the input signals can be tuned in order to control the microwave polarization at the center of the cross resonator in the first resonance, $f_d \sim 9.9$ GHz (at $T = 4$K). For signals of equal magnitude at ports $P_1$ and $P_2$, the relative phase, $\Delta\phi$, between them will determine the type of polarization obtained. For $\Delta\phi = m \times 180°$ (with $m = 0, \pm 1, \pm 2, \ldots$), the microwaves at the center of the resonator will be linearly polarized (coinciding with the destructive and constructive interferences of the quadrupole resonance), with the microwave magnetic field aligned ±45 degrees with respect to the resonator arms. For $\Delta\phi = n \times 90°$ (with $n = \pm 1, \pm 3, \pm 5, \ldots$), the microwaves will be circularly polarized. With $\Delta\phi = 90°$ and $270°$ giving rise to left ($\sigma^-$) and right ($\sigma^+$) circular polarizations, respectively, as indicated in Fig. 3.

To check the feasibility of our polarizable resonator we have recorded the EPR spectra of a single crystal ($\sim 0.3 \times 0.3 \times 0.3$ mm$^3$) of Fe$_{17}$Ga molecular wheels ($S = 5/2$) (see inset on Fig. 4a), diluted within a sea of antiferromagnetic Fe$_{18}$ wheels ($S = 0$), with a dilution factor 0.09:1 (Fe$_{17}$Ga:Fe$_{18}$). The purpose of working with a diluted sample is irrelevant in the scope of this work [17]. Nevertheless it provides a flavor of the high sensitivity of our resonator [18]. The Fe$_{17}$Ga molecules present a positive zero-field splitting, $DS_z^2$ ($D = 0.5$ K), making the $S_z = \pm 1/2$ spin projections the lowest energy states of the system. Experiments were carried out at 300 mK, where only the $S_z = \pm 1/2$ spin states are populated. Fig. 4a shows the EPR spectra of the sample obtained at 9.9 GHz for two different phase delays, $\Delta\phi = -90°$ ($\sigma^-$) and $90°$ ($\sigma^+$), between the microwave signals at ports $P_1$ and $P_2$. The observed EPR absorption peak at 0.32 T corresponds to a transition between the ground spin projection states ($S_z = \pm 1/2$) of the molecule ($g = 2$) [19]. The difference in the peak magnitude is associated to the selection rule imposed by spin transitions ($\Delta S = \pm 1$). In our case, the positive magnetic field makes the negative spin projection the ground state ($\Delta S = +1$), therefore, only right circular polarized microwaves ($\sigma^+$) are capable to induce transitions. Fig. 4b shows the modulation of the normalized height of the EPR peak as a function of the relative phase between the two microwave signals. Note that the EPR absorption peak corresponding to an $\Delta S = +1$ transition is proportional to the right polarized component of the microwave field. The continuous in Fig. 4b line represents the normalized magnitude of the $\sigma^+$-component of the microwave field, $\left(H_+ H_+^*\right)^{1/2} = (1+\sin\phi)^{1/2}$, with ideal control of the polarization angle. The results indicate that an 82% degree of polarization can be achieved within the volume of the sample [20]. Note that complete circular polarization is only achieved at the exact center of the cross, which has a total area of ~0.4×0.4 mm$^2$. Taking into account that the measured crystal covered more than 50% of the resonator area, the obtained degree of polarization is remarkable.

In conclusion, we have engineered a polarizable microstrip cross resonator that allows an arbitrary control of the polarization of high frequency microwaves without adjustable or moving mechanical actuators. The electronic control of the polarization provided by our microstrip cross resonator together with its low dimensions allows its integration in microscale devices for use in processes requiring fast modulation of the polarization (i.e. pulse EPR). In addition, the ability to measure in transmission mode allows for experiments at low temperatures, where long coaxial lines needed to wire the cryostat mask the reflected signals. Moreover, the idea of our microstrip cross resonator can be extended to *microstrip cross lines* or *coplanar cross waveguides* to generate circular polarized microwaves in a broad range of frequencies. Finally, our microstrip cross resonator provides a versatile tool for the study of fundamental characteristics of magnetic materials (i.e. molecular magnets), in where a selection of spin transitions can be realized. The ability to discriminate spin transitions becomes of crucial relevance for spin-based architectures in quantum information and computation applications.

**ACKNOWLEDMENTS**

We acknowledge fruitful discussions with Pieter Kik. E.d.B. acknowledges support from the US National Science Foundation (DMR0706183 and DMR0747587) and technical assistance from Agilent Technologies.

spins detected for a frequency bandwidth, $f$, of 1 Hz according to

$$N_{min} = N_{spins} / \left((S/N)\sqrt{f}\right).$$

[19] The position of the EPR peak (0.32 T) does not exactly match the expected resonant field of a $g = 2$ system at this frequency (~0.35T) due to the transverse anisotropy of the $Fe_{17}Ga$ molecules, which curves the spin projection levels.

[20] The degree of polarization has been calculated according to the next expression $P_+ = (A_{+90} - A_{-90})/(A_{+90} + A_{-90})$, where $A_{\Delta\phi}$ are the marginal values of the normalized heights of the EPR peaks, as shown in Fig. 4b.

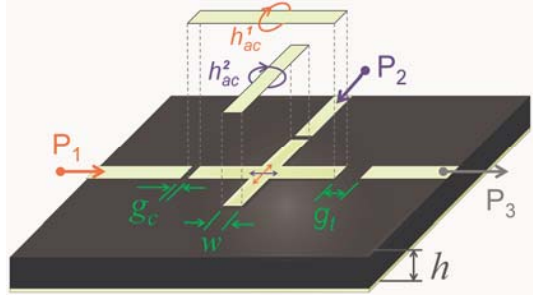

**FIG. 1:** (Color online) Polarizable *microstrip cross resonator*. Two half-wavelength microstrip line resonators are assembled together to form a cross resonator. The coupling, $g_c$, and transmission, $g_t$, gaps are engineered differently to optimize device performance. The relative magnitude and phase between ports $P_1$ and $P_2$ allow for controlling the polarization of microwaves at the center of the cross.

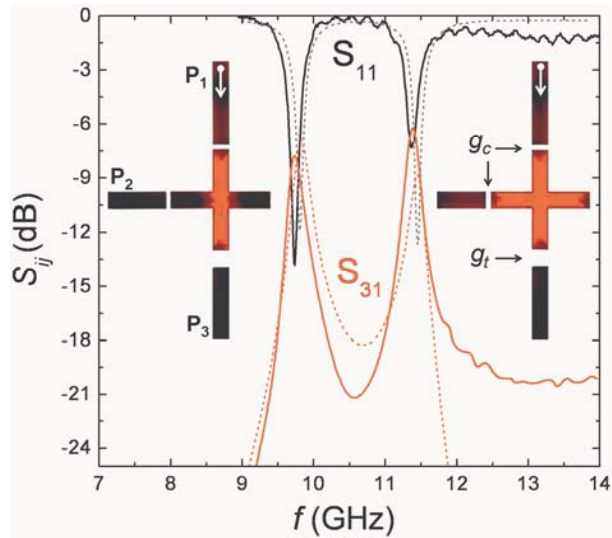

**FIG. 2:** (Color online) Reflection ($S_{11}$) and transmission ($S_{31}$) parameters measured on a 10 GHz microstrip cross resonator at room temperature. Two resonances are observed at 9.7 and 11.5 GHz. Dotted lines correspond to simulated data. The graphics show the current density at each of the resonances when microwave stimulus is applied only to port $P_1$.

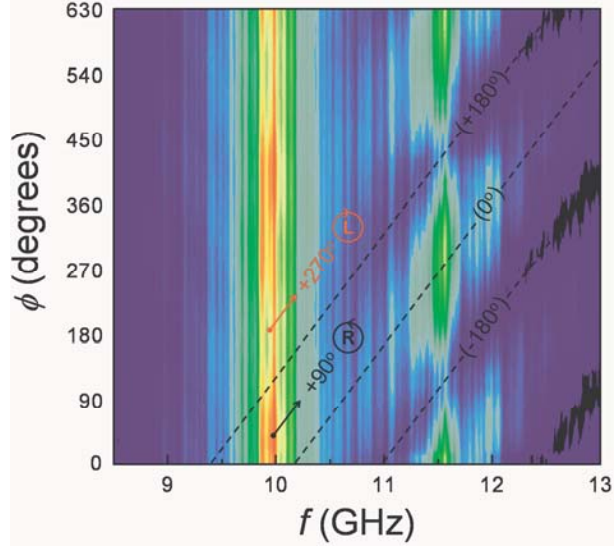

**FIG. 3:** (Color online) Contour plot of the transmission parameter ($S_{31}$) measured on a 10 GHz microstrip cross resonator at $T = 4$ K. The data shows the one-fold ($2\pi$) modulation of the resonances as a function of the relative phase between microwave signals of equal magnitude applied to ports $P_1$ and $P_2$. Dashed lines are used to relate the constructive (0°) and destructive (±180°) interferences of the quadrupole resonance (11.5 GHz) with the modulation observed in the dipolar resonance (~10 GHz). The slope of the lines is due to the linear frequency dependence of the phase supplied by the phase shifter employed.

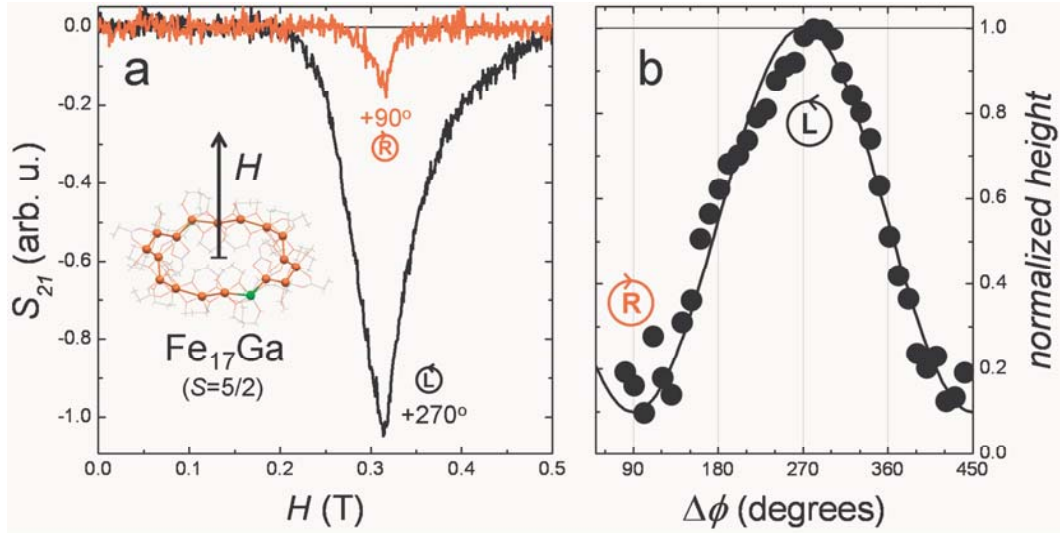

**FIG. 4:** (Color online) a) EPR spectra obtained at 300 mK on a magnetically dilute single crystal of $Fe_{17}Ga$ molecular magnets. The two curves correspond to two different phase delays, $\Delta\phi = -90°$ ($\sigma^-$) and $90°$ ($\sigma^+$), between the microwave signals at ports $P_1$ and $P_2$ of the cross resonator. The inset shows and sketch of the $Fe_{17}Ga$ molecular wheel, in which one Fe ($S = 5/2$) ion has been supplanted by a Ga ($S = 0$), resulting in a frustrated molecular spin $S = 5/2$ at low temperatures. b) Modulation of the normalized EPR peak height as a function of the phase delay between the signals. A 82% degree of circular polarization is achieved over the central area of the resonator.